\documentclass[conference]{IEEEtran}
\usepackage[a4paper, total={7.3in, 9.8in}]{geometry}
\IEEEoverridecommandlockouts
\usepackage{amsmath,amssymb,amsfonts}
\usepackage{algorithmic}
\usepackage{graphicx}
\usepackage{textcomp}
\def\BibTeX{{\rm B\kern-.05em{\sc i\kern-.025em b}\kern-.08em
    T\kern-.1667em\lower.7ex\hbox{E}\kern-.125emX}}
\usepackage[utf8]{inputenc}
\usepackage{datetime}
\usepackage{upgreek}
\usepackage{dblfloatfix}    
\usepackage{subfigure}
\renewcommand{\figurename}{Fig.}
\usepackage{multirow}
\usepackage{multicol}
\usepackage{booktabs}
\usepackage{caption}
\usepackage{subcaption}
\usepackage{graphicx}
\usepackage{soul,lipsum}
\usepackage{siunitx}
\usepackage{xspace}
\usepackage{bm}
\usepackage{makecell} 
\usepackage{xr-hyper}
\usepackage{hyperref}
\hypersetup{
    colorlinks=true,
    linkcolor=blue,
    filecolor=magenta,      
    urlcolor=cyan,
}
\usepackage[capitalize]{cleveref}
\usepackage{etoolbox}
\patchcmd{\thebibliography}{\section*{\refname}}{}{}{}
\usepackage{pifont} 

\def\BibTeX{{\rm B\kern-.05em{\sc i\kern-.025em b}\kern-.08em
    T\kern-.1667em\lower.7ex\hbox{E}\kern-.125emX}}

\usepackage{threeparttable}
\makeatother
\usepackage[hyphenbreaks]{breakurl}
\makeatletter
\usepackage[style=ieee, backend=biber, defernumbers=false, uniquename=false, uniquelist=false, maxnames=1, maxbibnames=1, minnames=1]{biblatex}
\AtEveryBibitem{%
  \clearfield{title}   
}

\usepackage{upgreek} 

\usepackage[dvipsnames,svgnames]{xcolor} 

\usepackage[draft]{pdfcomment} 
\usepackage[hyperfirst=false,acronym,sort=none,shortcuts,nopostdot,style=super,nonumberlist,toc,nogroupskip]{glossaries}

\glsdisablehyper 

\newcommand{\accolor}[1]{\textcolor{Black}{#1}}

\newacronymstyle{myacro}
{%
  \GlsUseAcrEntryDispStyle{long-short}%
}%
{%
  \GlsUseAcrStyleDefs{long-short}%
}

\setacronymstyle{myacro}

\newcommand*{\tip}[1]{
    \ifglsused{#1}{
      {\pdftooltip{\accolor{\glsentryshort{#1}}}{\glsentrydesc{#1}}}%
    }{%
      \gls{#1}
    }%
}%

\newcommand*{\tips}[1]{
    \ifglsused{#1}{
      {\pdftooltip{\accolor{\glsentryshortpl{#1}}}{\glsentrydescplural{#1}}}%
    }{%
      \glspl{#1}
    }%
}%

\newcommand{\xx}[1]{\tip{#1}}

\newacronym{DVS}{DVS}{Dynamic Vision Sensor}
\newacronym{dvs}{DVS}{Dynamic Vision Sensor}
\newacronym{PSD}{PSD}{Power Spectral Density}
\newacronym{TC}{TC}{Temporal Contrast}
\newacronym{SNR}{SNR}{Signal-to-Noise Ratio}
\newacronym{RMS}{RMS}{root mean square}
\newacronym{SF}{SF}{Source-Follower buffer}
\newacronym{theta_on}{$\Theta_{\text{ON}}$}{ON threshold}
\newacronym{theta_off}{$\Theta_{\text{OFF}}$}{OFF threshold}

\newacronym[description={False Negative Rate; signal that is incorrectly classified as noise}]{fnr}{FNR}{False Negative Rate}
\newacronym[description={False Negative; signal event that is incorrectly classified as noise event}]{fn}{FN}{False Negative}
\newacronym[description={False Positive Rate; noise that is incorrectly classified as signal}]{fpr}{FPR}{False Positive Rate}
\newacronym[description={False Positive; noise event that is incorrectly classified as signal event}]{fp}{FP}{False Positive}
\newacronym[description={Guided Event Filtering: Joint Filtering of Intensity Images and Neuromorphic Events}]{gef}{GEF}{Guided Event Flow}
\newacronym[description={Inter Spike Interval (nomenclature from neuroscience)}]{isi}{ISI}{Inter Spike Interval}
\newacronym[description={MAC (Multiply-Accumulate) is the basic operation of signal processing and artificial neural networks. One MAC is 2 Op.}]{mac}{MAC}{Multiply-Accumulate}
\newacronym[description={Multipurpose block random access memory module in FPGA}]{bram}{BRAM}{Block RAM}
\newacronym[description={Register Transfer Logic intermediate form, consisting of combinational and synchronous register logic cells}]{rtl}{RTL}{Register Transfer Logic}
\newacronym[description={Single Threshold Metric; a measure of the ROC TPR/FPR tradeoff at one discrimination threshold}]{stm}{STM}{Single Threshold Metric}
\newacronym[description={Surface of Active Event; image of latest event timestamps, same as Timestamp Image}]{sae}{SAE}{Surface of Active Events}
\newacronym[description={System on Chip; FPGA with embedded programmable processor}]{soc}{SoC}{System on Chip}
\newacronym[description={Time Surface; image of age of events relative to a particular event}]{ts}{TS}{Time Surface}
\newacronym[description={Timestamp Image; 2D image of latest event timstamps, similar to Surface of Active Events}]{ti}{TI}{Timestamp Image}
\newacronym[description={Timestamp+Polarity Image; 2D image of latest event timstamps and +/- brightness change polarites}]{tpi}{TPI}{Timestamp+Polarity Image}
\newacronym[description={True Negative Rate; noise that is correctly classified as noise}]{tnr}{TNR}{True Negative Rate}
\newacronym[description={True Negative; noise that is correctly classified as noise}]{tn}{TN}{True Negative}
\newacronym[description={True Positive Rate; signal that is correctly classified as signal}]{tpr}{TPR}{True Positive Rate}
\newacronym[description={True Positive; signal event that is correctly classified as signal}]{tp}{TP}{True Positive}
\newacronym[longplural={Convolutional Neural Networks}]{cnn}{CNN}{Convolutional Neural Network}
\newacronym[longplural={First In First Out memories}]{fifo}{FIFO}{First In First Out memory}
\newacronym{adc}{ADC}{Analog to Digital Converter}
\newacronym{aer}{AER}{Address Event Protocol}
\newacronym{aps}{APS}{Active Pixel Sensor}
\newacronym{asic}{ASIC}{Application Specific Integrated Circuit}
\newacronym{auc}{AUC}{Area Under the Curve}
\newacronym{baf}{BAF}{Background Activity Filter}
\newacronym{ba}{BA}{Background Activity}
\newacronym{bmof}{BMOF}{Block Matching Optical Flow}
\newacronym{bm}{BM}{Block Matching}
\newacronym{bp}{BP}{Back Propagation}
\newacronym{cfa}{CFA}{Color Filter Array}
\newacronym{cf}{CF}{Complementary Filter}
\newacronym{cg}{CG}{Convolutional Gated Recurrent Unit Network}
\newacronym{cis}{CIS}{CMOS Image Sensor}
\newacronym{cmae}{CMAE}{Cross-Modality Attention Enhancement}
\newacronym{contrastmaximization}{CM}{Contrast Maximization}
\newacronym{cots}{COTS}{Commodity Off-The-Shelf}
\newacronym{cpu}{CPU}{Central Processing Unit}
\newacronym{cv}{CV}{Computer Vision}
\newacronym{davis}{DAVIS}{Dynamic and Active pixel Vision Sensor}
\newacronym{dba}{DBA}{Dynamic Background Activity noise filtering algorithm}
\newacronym{dnn}{DNN}{Deep Neural Network}
\newacronym{dof}{DOF}{Degree of Freedom}
\newacronym{dolp}{DoLP}{Degree of Linear Polarization}
\newacronym{dram}{DRAM}{Dynamic RAM}
\newacronym{drcn}{DRCN}{Deep Recurrent Convolutional Network}
\newacronym{dr}{DR}{Dynamic Range}
\newacronym{dsp}{DSP}{Digital Signal Processing unit}
\newacronym{dwf}{DWF}{Double Window Filter}
\newacronym{e2pd}{E2PD}{Events to Polarization Dataset}
\newacronym{e2p}{E2P}{Events to Polarization}
\newacronym{edflow}{EDFLOW}{Event-driven Optical Flow}
\newacronym{edncnn}{EDnCNN}{Event Denoising CNN}
\newacronym{edp}{EDP}{Event Denoising Precision}
\newacronym{efast}{EFAST}{Event-Based time surface FAST}
\newacronym{epm}{EPM}{Event Probability Mask}
\newacronym{fast}{FAST}{Features from Accelerated Segment Test}
\newacronym{feast}{FEAST}{Feature Extraction with Adaptive Selection Thresholds }
\newacronym{flipflop}{FF}{Flip-Flop}
\newacronym{fom}{FOM}{Figure of Merit}
\newacronym{fpga}{FPGA}{Field Programmable Gate Array}
\newacronym{fpn}{FPN}{Fixed Pattern Noise}
\newacronym{fps}{FPS}{Frames Per Second}
\newacronym{fsae}{FSAE}{Filtered Surface of Active Events}
\newacronym{fwf}{FWF}{Fixed Window Filter}
\newacronym{gpu}{GPU}{Graphics Processing Unit}
\newacronym{gt}{GT}{Ground Truth}
\newacronym{hdl}{HDL}{Hardware Description Language}
\newacronym{hdr}{HDR}{high dynamic range}
\newacronym{hls}{HLS}{High Level Synthesis}
\newacronym{icm}{ICM}{Iterated Conditional Modes}
\newacronym{id}{ID}{Index Decay}
\newacronym{iir}{IIR}{Infinite Impulse Response}
\newacronym{imu}{IMU}{Inertial Measurement Unit}
\newacronym{inceptiveevent}{IE}{Inceptive Event}
\newacronym{iot}{IoT}{Internet of Things}
\newacronym{ip}{IP}{Intellectual Property}
\newacronym{its}{ITS}{Invariant Time Surface}
\newacronym{knn}{KNN}{$K$-Nearest-Neighbor clustering}
\newacronym{li}{LI}{Leaky Integrator}
\newacronym{lk}{LK}{Lucas-Kanade}
\newacronym{lpips}{LPIPS}{Learned Perceptual Image Patch Similarity}
\newacronym{lut}{LUT}{LookUp Table}
\newacronym{mlpf}{MLPF}{MultiLayer Perceptron denoising Filter}
\newacronym{mlp}{MLP}{Multilayer Perceptron}
\newacronym{ml}{ML}{Machine Learning}
\newacronym{mpeg}{MPEG}{Motion Picture Experts Group}
\newacronym{mse}{MSE}{Mean Squared Error}
\newacronym{na}{NA}{Numerical Aperture}
\newacronym{nnb}{NNb}{Nearest Neighbor}
\newacronym{of}{OF}{Optical Flow}
\newacronym{onf}{ONF}{Order(N) Filter}
\newacronym{pcb}{PCB}{Printed Circuit Board}
\newacronym{pdavis}{PDAVIS}{Polarization Dynamic and Active pixel VIsion Sensor}
\newacronym{pd}{PD}{photodiode}
\newacronym{pe}{PE}{Processing Element}
\newacronym{pfa}{PFA}{Polarization Filter Array}
\newacronym{pl}{PL}{programmable Logic}
\newacronym{por}{POR}{Positive Output Ratio}
\newacronym{prm}{PRM}{Pixel Rendering Module}
\newacronym{ps}{PS}{Processing System}
\newacronym{pugm}{PUGM}{Probabilistic Undirected Graph Model}
\newacronym{qwp}{QWP}{Quarter Wave Plate}
\newacronym{ram}{RAM}{Random Access Memory}
\newacronym{ransac}{RANSAC}{Random Sample and Consensus}
\newacronym{ratp}{RATP}{Recursive Adaptive Temporal Pooling}
\newacronym{rb}{RB}{Residual Block}
\newacronym{relu}{ReLU}{Rectified Linear Unit}
\newacronym{roc}{ROC}{Receiver Operating Characteristic}
\newacronym{roi}{ROI}{Region of Interest}
\newacronym{rpmd}{RPMD}{Relative Plausibility Measure of Denoising}
\newacronym{rpm}{RPM}{Revolutions per Minute}
\newacronym{rppp}{RPPP}{Rich Polarization Pattern Perception}
\newacronym{sad}{SAD}{Sum of Absolute Differences}
\newacronym{sm}{SM}{Supplementary Material}
\newacronym{snr}{SNR}{Signal to Noise Ratio}
\newacronym{soa}{SOA}{state of the art}
\newacronym{sram}{SRAM}{Static RAM}
\newacronym{stcf}{STCF}{SpatioTemporal Correlation Filter}
\newacronym{tda}{TDA}{Time Decay Adapted}
\newacronym{td}{TD}{Time Decay}
\newacronym{timsl}{TS}{time slice}
\newacronym{usb}{USB}{Universal Serial Bus}
\newacronym{vga}{VGA}{Video Graphics Adaptor}
\newacronym{vhdl}{VHDL}{Very High-Speed Integrated Circuit Hardware Description Language}
\newacronym{zoh}{ZOH}{Zero-Order Hold}

\newcommand*{\Ipd}{\ensuremath{I_\text{pd}}\xspace}
\newcommand*{\Ipr}{\ensuremath{I_\text{pr}}\xspace}

\newcommand*{\Vpd}{\ensuremath{V_\text{pd}}\xspace}
\newcommand*{\Vpr}{\ensuremath{V_\text{pr}}\xspace}

\newcommand*{\Mampn}{\ensuremath{M_\text{amp,n}}\xspace}
\newcommand*{\Mampp}{\ensuremath{M_\text{amp,p}}\xspace}

\newcommand*{\Vdiff}{\ensuremath{V_\text{diff}}\xspace}
\newcommand*{\Vsf}{\ensuremath{V_\text{sf}}\xspace}

\newcommand*{\Isf}{\ensuremath{I_\text{sf}}\xspace}

\newcommand*{\Ut}{\ensuremath{U_\text{T}}\xspace}

\newcommand*{\Cpd}{\ensuremath{C_\text{pd}}\xspace}
\newcommand*{\Cpr}{\ensuremath{C_\text{pr}}\xspace}
\newcommand*{\Cfb}{\ensuremath{C_\text{fb}}\xspace}
\newcommand*{\Csf}{\ensuremath{C_\text{sf}}\xspace}

\newcommand*{\Rout}{\ensuremath{R_\text{out}}\xspace}

\newcommand*{\gmfb}{\ensuremath{g_{\text{m}_\text{fb}}}\xspace}
\newcommand*{\gm}{\ensuremath{g_{\text{m}}}\xspace}
\newcommand*{\gs}{\ensuremath{g_{\text{s}}}\xspace}
\newcommand*{\gsfb}{\ensuremath{g_{\text{s}_\text{fb}}}\xspace}
\newcommand*{\gssf}{\ensuremath{g_{\text{s}_\text{sf}}}\xspace}

\newcommand*{\gmamp}{\ensuremath{g_{\text{m}_\text{amp,n}}}\xspace}
\newcommand*{\Aloop}{\ensuremath{A_\text{loop}}\xspace}
\newcommand*{\Zmdc}{\ensuremath{Z_{\text{m}_\text{dc}}}\xspace}
\newcommand*{\wzzm}{\ensuremath{w_{\text{z}_{\text{Z}_\text{m}}}}\xspace}

\newcommand*{\Zoutdc}{\ensuremath{Z_{\text{out}_\text{dc}}}\xspace}
\newcommand*{\wzzout}{\ensuremath{w_{\text{z}_{\text{Z}_\text{out}}}}\xspace}

\newcommand*{\Gmfb}{\ensuremath{G_{\text{m}_\text{fb}}}\xspace}

\newcommand*{\Zms}{\ensuremath{Z_{\text{m}}(s)}\xspace}
\newcommand*{\Zouts}{\ensuremath{Z_{\text{out}}(s)}\xspace}

\newcommand*{\Zmsqf}{\ensuremath{Z_{\text{m}}^2(f)}\xspace}
\newcommand*{\Zoutsqf}{\ensuremath{Z_{\text{out}}^2(f)}\xspace}
\newcommand*{\taupd}{\ensuremath{\tau_\text{pd}}\xspace}
\newcommand*{\tausf}{\ensuremath{\tau_\text{sf}}\xspace}
\newcommand*{\taupr}{\ensuremath{\tau_\text{pr}}\xspace}

\newcommand*{\Asfsqf}{\ensuremath{A_{\text{sf}}^2(f)}\xspace}
\newcommand*{\Zoutsfsqf}{\ensuremath{Z_{\text{out}_\text{sf}}^2(f)}\xspace}

\newcommand*{\kappafb}{\ensuremath{\kappa_{\text{fb}}}\xspace}
\newcommand*{\kappaampn}{\ensuremath{\kappa_{\text{amp,n}}}\xspace}
\newcommand*{\Vaampn}{\ensuremath{V_{\text{A}_\text{amp,n}}}\xspace}
\newcommand*{\Vaampp}{\ensuremath{V_{\text{A}_\text{amp,p}}}\xspace}
\newcommand*{\kappasf}{\ensuremath{\kappa_{\text{sf}}}\xspace}

\newcommand*{\Zdc}{\ensuremath{\text{Z}_\text{dc}}\xspace}
\newcommand*{\Zm}{\ensuremath{\text{Z}_\text{m}}\xspace}
\newcommand*{\Zout}{\ensuremath{\text{Z}_\text{out}}\xspace}

\newcommand*{\Ipds}{\ensuremath{\text{I}_\text{pd}\text{(s)}}\xspace}
\newcommand*{\Iprs}{\ensuremath{\text{I}_\text{pr}\text{(s)}}\xspace}
\newcommand*{\Vprs}{\ensuremath{\text{V}_\text{pr}\text{(s)}}\xspace}
\newcommand*{\Vsfs}{\ensuremath{\text{V}_\text{sf}\text{(s)}}\xspace}
\newcommand*{\Isfs}{\ensuremath{\text{I}_\text{sf}\text{(s)}}\xspace}
\newcommand*{\Asfs}{\ensuremath{\text{A}_\text{sf}\text{(s)}}\xspace}
\newcommand*{\Zsfs}{\ensuremath{\text{Z}_{\text{out}_\text{sf}}\text{(s)}}\xspace}

\begin{document}

\title{Towards a physically realistic \\ computationally efficient DVS pixel model}

\author{\IEEEauthorblockN{Rui Graca, Tobi Delbruck}
\IEEEauthorblockA{\textit{Sensors Group, Inst. of Neuroinformatics, UZH-ETH Zurich, 
Zurich, Switzerland} \\
rpgraca,tobi@ini.ethz.ch, \url{https://sensors.ini.ch}}
}

\maketitle



\begin{abstract}
Dynamic Vision Sensor (DVS) event camera models are important tools for predicting camera response, optimizing biases, and generating realistic simulated datasets. Existing DVS models have been useful, but have not demonstrated high realism for challenging HDR scenes combined with adequate computational efficiency for array-level scene simulation. This paper reports progress towards a physically realistic and computationally efficient DVS model based on large-signal differential equations derived from circuit analysis, with parameters fitted from pixel measurements and circuit simulation. These are combined with an efficient stochastic event generation mechanism based on first-passage-time theory, allowing accurate noise generation with timesteps greater than 1000x longer than previous methods. 
\end{abstract}

\section{Introduction}
\label{sec:intro}
\tip{DVS} event cameras \cite{lichtsteiner2008latencyasynchronous,gallego2019eventbased}
are gaining popularity due to their high-speed, sparse output, low-power, and high dynamic range.
\tip{DVS} encode brightness changes in a visual scene into a stream of ON and OFF events, resulting in an output data rate determined by the activity in the scene.

\tips{DVS} have applications in areas such as robotics and image deblurring~\cite{gallego2019eventbased}. \tip{DVS}  pixel sizes have reduced following technology scaling and wafer stacking~\cite{guo2023waferstacked,kodama2023rgbhybrid,niwa2023eventbased}. Other \gls{DVS} implementations focus on reducing the minimum achievable event threshold for on scientific applications \cite{serrano2013contrastsensitivity,moeys2018sensitivedynamic,graca2024scidvs}

\tip{DVS} introduces new challenges, such as: 1. the \tip{DVS} pixel is significantly more complex than the APS, with operation controlled by several user-defined bias currents, therefore requiring more careful optimization; and 2. there is a lack of \tip {DVS} datasets in comparison to frame-based datasets.

Both of these challenges can be mitigated by pixel models: Realistic models would allow the user to optimize bias settings offline, as well as to generate simulated datasets from either frame-based datasets or descriptions of the scene. 

\tip{DVS} models including the popular ESIM and v2e focus on efficiently generating events from frame-based videos for deep learning datasets~\cite{rebecq2018esim,hu2021v2e,lin2022dvsvoltmeter,zhang2024v2ce,jiang2024adv2e}. They fall short in realistically modeling the dependence of the \tip{DVS} on illuminance or bias settings regarding the dynamics of both signal and noise. 
Existing models are too inaccurate to be used as a tool for offline bias optimization or to generate artificial training data in challenging conditions of extreme dynamic range and high noise, such as those encountered in space domain awareness and some scientific imaging.

More physically realistic \tip{DVS} models use differential equations derived from circuit analysis to match SPICE simulations and measured event threshold variation \cite{suess2023physicalmodeling,mou2024jointparameter,Joubert2021-event-simulator-improvements}, resulting in a more accurate dependence on illuminance and bias settings.
However, accurately modeling noise requires tiny timesteps, since the model will ignore events happening due to transitory crossings of the threshold between consecutive timesteps, leading to severe underprediction of the effect of noise if the simulation timestep is too long, which makes the model too slow for the generation of array-level video data. 
Besides, model generalization across bias settings was not demonstrated, making them unsuitable for offline bias optimization.

\begin{figure}[h]
    \centering
    \includegraphics[width=0.8\columnwidth]{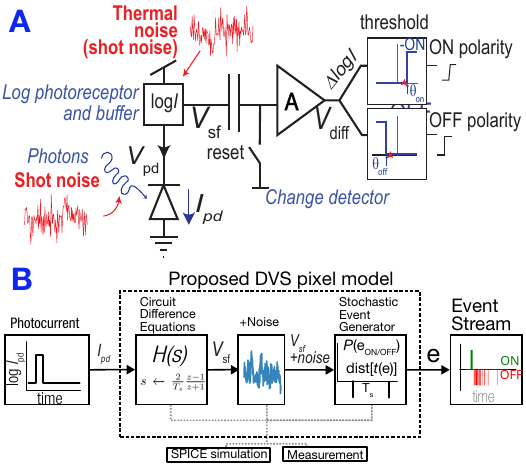}
    \caption{\textbf{A:} DVS pixel architecture. \textbf{B:} Diagram of the proposed \xx{dvs} pixel model with input \Ipd and output events \textsf{e}.}
    \label{fig:model_diagram}
\end{figure}

\subsection{Contributions}
\cref{fig:model_diagram} shows an overview of the \xx{dvs} pixel (\textbf{A}) our the proposed \xx{dvs} pixel model (\textbf{B}).
\begin{enumerate}
    \item The model starts with 
circuit analysis (\cref{fig:schematic}), resulting in a second-order differential equation for the photoreceptor and a first-order differential equation for the buffer \cite{graca2023optimalbiasing,graca2023shininglight}. The small signal model parameters are updated at each timestep for large signal modeling. 
\item For event generation, besides simply assessing if the threshold was crossed at the current timestep \cite{hu2021v2e}, our model calculates the probability of noise causing a transitory crossing of the threshold in the interval between timesteps to stochastically generate events with physically accurate timing. This method (\cref{fig:model_diagram}\textbf{B}) allows increasing the simulation timestep by a factor of 3 orders of magnitude without loss of accuracy. 
\item  The model parameters are physically meaningful circuit parameters determined from bias currents, capacitance values, and transistor parameters, validated by pixel measurements and SPICE simulations~\cite{graca2024eventcamera}.
\end{enumerate}

The remainder of the paper is organized as follows:
\cref{sec:prmodel} introduces the photoreceptor model and buffer models, \cref{sec:fitting} describes how model parameters are fitted from measurements, \cref{sec:tran} presents results from large-signal temporal simulation, and \cref{sec:eventgen} describes how events are stochastically generated between timesteps.

\section{Photoreceptor Model}
\label{sec:prmodel}
\begin{figure}[tb]
    \centering
    \includegraphics[width=0.3\textwidth]{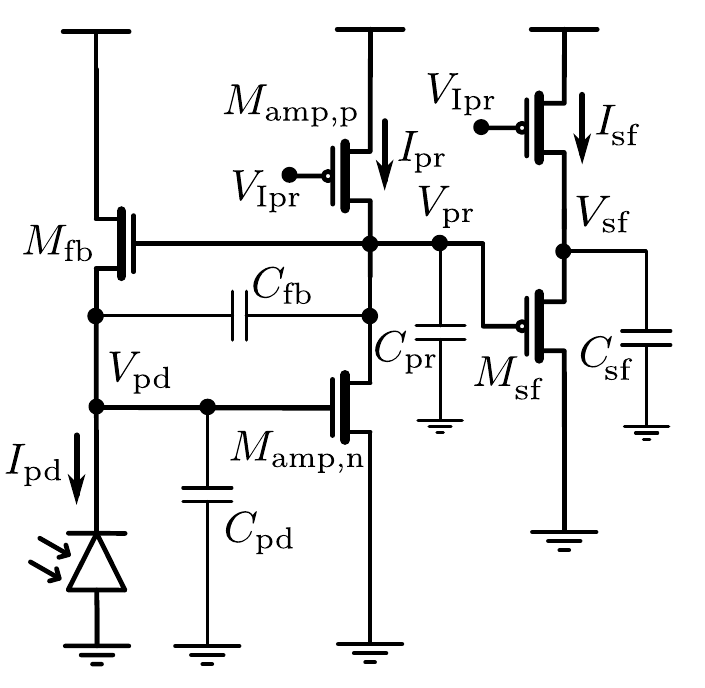}
    \caption{Photoreceptor (PR) and source-follower (SF) buffer schematic.}
    \label{fig:schematic}
\end{figure}

Referring to \cref{fig:schematic}, our photoreceptor and buffer model consists of four transfer functions derived from circuit analysis: $\Zms=\frac{\Vprs}{\Ipds}$, $\Zouts=\frac{\Vprs}{\Iprs}$, $\Asfs=\frac{\Vsfs}{\Vprs}$, and $\Zsfs=\frac{\Vsfs}{\Isfs}$. \Zm and \Zout are second-order, given by:  
\begin{equation}
  \label{eq:zm_sfpr}
  Z(s) = \frac{\Zdc(1+\frac{s}{w_z})}{\frac{s^2}{w_0^2}+2\frac{\zeta}{w_0}s+1}
\end{equation}

\noindent where $\wzzm=-\frac{\gmamp}{\Cfb}$ and $\wzzout=-\frac{\gsfb}{\Cpd+\Cfb}$, $\Zmdc = \frac{1}{\gmfb}\frac{\Aloop}{\Aloop+1}$, $\Zoutdc = \frac{\Rout}{\Aloop+1}$, $w_0^2=\frac{(\Cpd\Cfb+\Cpr\Cfb+\Cpr\Cpd)\frac{\Rout}{\gsfb}}{\Aloop+1}$, $2\frac{\zeta}{w_0}=\frac{\frac{1}{\gsfb}(\Cpd+(1+\gmamp\Rout)\Cfb)+\Rout(\Cpr+(1-\frac{\Gmfb}{\gsfb})\Cfb)}{\Aloop+1}$, and $\Aloop=\frac{\gmamp\Rout\gmfb}{\gsfb}$. 

Assuming weak inversion for all transistors, $\gm=\frac{\kappa I}{\Ut}$ and $\gs=\frac{I}{\Ut}$. $\Rout=\frac{\Vaampn\Vaampp}{\Ipr(\Vaampn+\Vaampp)}$ is the parallel resistance between the output resistances of \Mampn and \Mampp, where \Vaampn and \Vaampp are the Early voltages of \Mampn and \Mampp. The derivation of these equations can be found in~\cite{graca2024eventcamera}.

If the \Vpd time constant $\taupd=\frac{1}{\gsfb}(\Cpd+(1+\gmamp\Rout)\Cfb)$ is much larger than the \Vpr time constant $\taupr=\Rout(\Cpr+(1-\frac{\Gmfb}{\gsfb})\Cfb)$, which is generally the case when $\Ipr\gg\Ipd$, \taupd becomes dominant, leading to the results in \cite{suess2023physicalmodeling}.

\Asfs and \Zsfs are given by:
\vspace{-3em} 
\begin{multicols}{2}
  \begin{equation}
    \Asfs = \frac{\kappa_\text{sf}}{1+s\frac{\Csf}{\gssf}}
  \end{equation}\break
  \begin{equation}
    \Zsfs = \frac{\frac{1}{\gssf}}{1+s\frac{\Csf}{\gssf}}
  \end{equation}
\end{multicols}

The signal transfer function is given by $\frac{\Vsfs}{\Ipds}=\Zms\Asfs$.

\textbf{Noise:} The shot noise \gls{PSD} at \Vsf as described in \cite{graca2023optimalbiasing,suess2023physicalmodeling} is given by the sum of the contributions of \Ipd, \Ipr, and \Isf, respectively: $4q\Ipd\Zmsqf\Asfsqf$, $4q\Ipr\Zoutsqf\Asfsqf$, and $4q\Isf\Zoutsfsqf$.

\section{Parameter fitting}
\label{sec:fitting}
\begin{figure*}[h]
    \centering
    \includegraphics[width=0.8\textwidth]{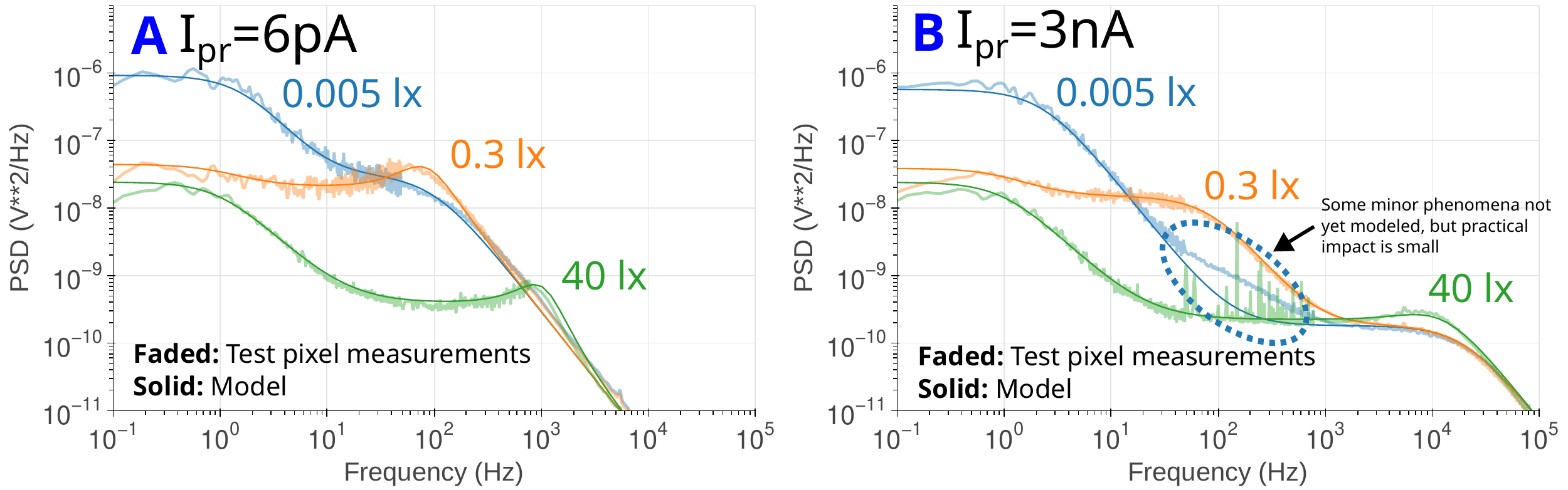}
    \caption{Measured pixel PSD (faded lines) and proposed model (solid lines) for different on-chip illuminances for (\textbf{A}) \Ipr of \qty{6}{\pico\ampere} and (\textbf{B}) \Ipr of \qty{3}{\nano\ampere}.}
    \label{fig:psd_fit}
\end{figure*}

The model described in \cref{sec:prmodel} depends on the capacitances \Cpd, \Cfb, \Cpr, and \Csf, currents \Ipd, \Ipr, and \Isf, and transistor parameters \kappafb, \kappaampn, \kappasf, \Vaampn, and \Vaampp. These parameters can be estimated from SPICE simulations, and fine-tuned to match pixel measurements. \cref{fig:psd_fit} shows how the model matches noise \gls{PSD} curves measured from a DAVIS346 test pixel. Flicker noise was added to match the low frequency response in bright conditions. However, the practical impact of flicker is generally very small and it can be neglected. The curves show that our model generalizes well for different illuminations across different bias settings. 

\section{Large-signal temporal simulation}
\label{sec:tran}
To simulate the temporal response of the \xx{DVS} pixel, we discretize the differential equations using a bilinear transform $s\leftarrow\frac{2}{T_s}\frac{z-1}{z+1}$, where $T_s$ is the simulation timestep, and solve the resulting difference equations.  The large-signal response is approximated by linearizing the system at each timestep.

\textbf{Noise:} Randomly generated Gaussian noise as described in \cref{sec:prmodel} is filtered separately and added to \Vpr. Then \Vpr is filtered by the source-follower buffer difference equations and the SF noise is filtered separately and added to \Vsf.

\begin{figure*}[h]
    \centering
    \includegraphics[width=.75\textwidth]{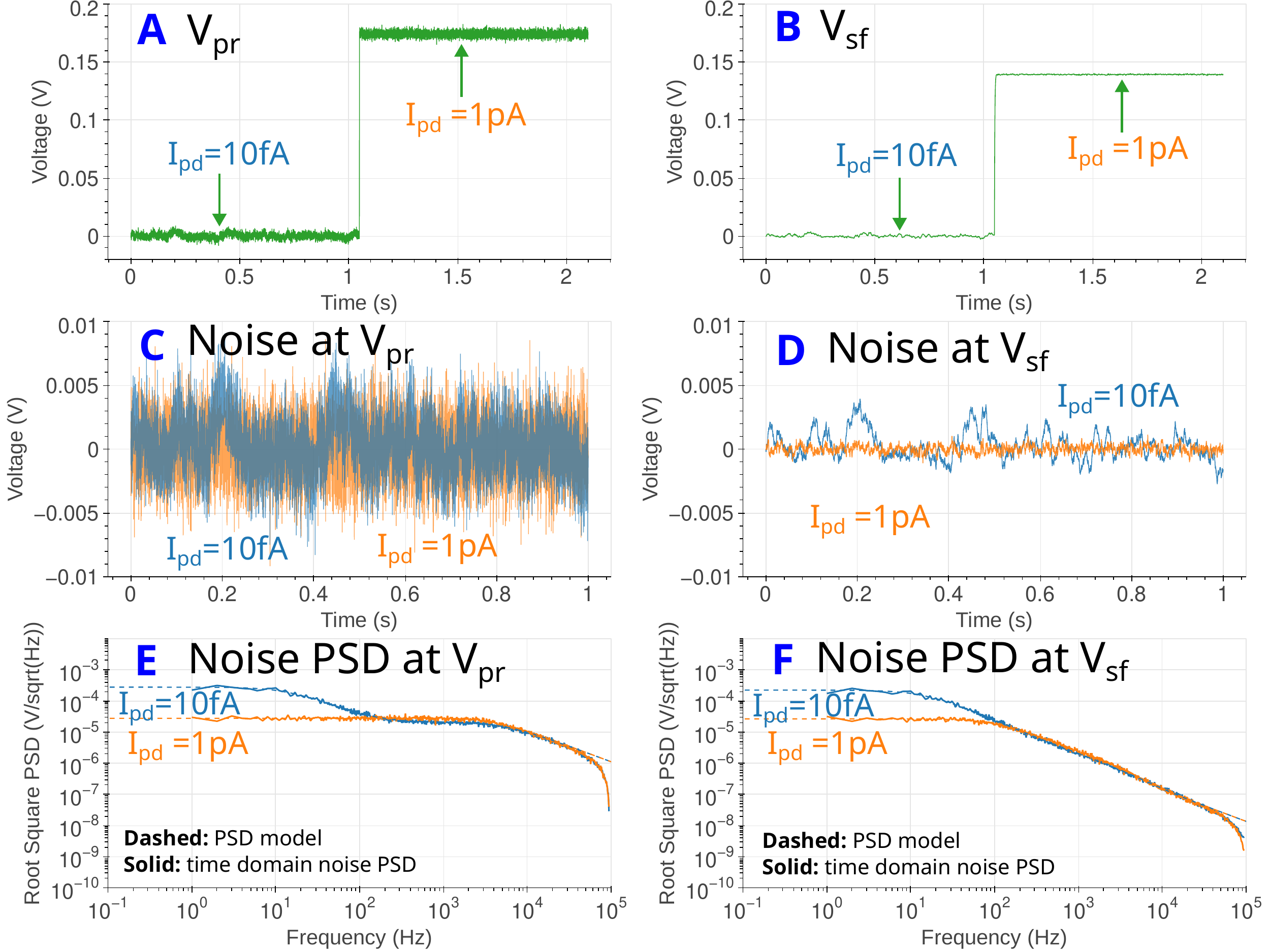}
    \caption{Model response to a step in photocurrent \Ipd between \qty{10}{\femto\ampere} and \qty{1}{\pico\ampere}. (\textbf{A}) shows the temporal response at \Vpr and (\textbf{B}) at \Vsf. (\textbf{C}) shows the overlapped zoomed-in noise at for the different values of \Ipd at \Vpr, and (\textbf{D}) at \Vsf. (\textbf{E}) and (\textbf{F}) show the PSD from (\textbf{D}) and (\textbf{E}) (solid), as well as the fitted PSD model for the same conditions (dashed).}
    \label{fig:large_signal_time_domain_noise}
\end{figure*}

\cref{fig:large_signal_time_domain_noise} shows the transient simulation of a 100x large-signal step in \Ipd from \qty{10}{\femto\ampere} to \qty{1}{\pico\ampere}. 
The simulation timestep was \qty{10}{\micro\second}.
\cref{fig:large_signal_time_domain_noise}\textbf{A}\&\textbf{B} shows the simulated voltage response at \Vpr and \Vsf. 
\cref{fig:large_signal_time_domain_noise}\textbf{C}\&\textbf{D} shows the steady-state temporal voltage noise at \Vpr and \Vsf (respectively) for both values of \Ipd.
\cref{fig:large_signal_time_domain_noise}\textbf{E}\&\textbf{F} shows their \xx{PSD}, which accurately match the theoretical \xx{PSD} model fitted from pixel measurements. This simulation shows our model can generalize and produce physically realistic noise at different photocurrent levels for a large-signal simulation. 

\begin{figure*}[h]
    \centering
    \includegraphics[width=.75\textwidth]{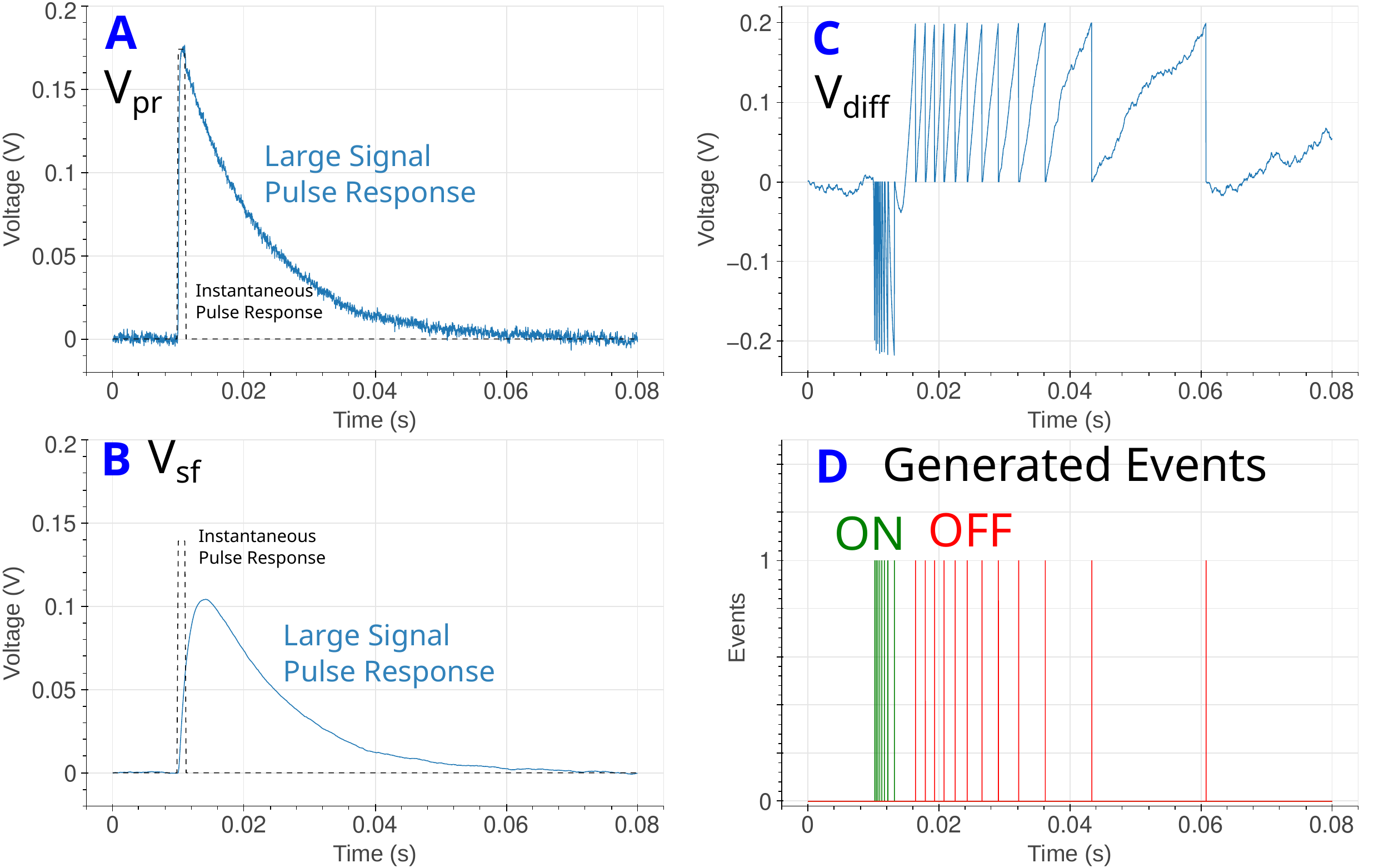}
    \caption{Model response to a \qty{1}{\milli\second} pulse from \qty{10}{\femto\ampere} to \qty{1}{\pico\ampere} in photocurrent. (\textbf{A}) shows simulated signal at \Vpr, (\textbf{B}) at \Vsf, and (\textbf{C}) at \Vdiff. (\textbf{D}) shows the generated ON and OFF events.}
    \label{fig:large_signal_time_domain_pulse}
\end{figure*}

\cref{fig:large_signal_time_domain_pulse} shows the model's response to a \qty{1}{\milli\second} pulse in \Ipd from \qty{10}{\femto\ampere} to \qty{1}{\pico\ampere}. The model successfully reproduces the asymmetrical non-linear behavior observed in practice, where the falling edge of the pulse results in much slower response than the rising edge, leading to a trail of OFF events.
Noise is most clearly visible in \cref{fig:large_signal_time_domain_pulse}C.

\section{Stochastic event generation}
\label{sec:eventgen}
Generating events only when the threshold is crossed at a given time step can lead to a severe underestimation of the effect of noise. Noise can cause momentary threshold crossings anytime between timesteps, and the fact that \Vdiff is below threshold at consecutive timesteps does not guarantee that it remained below threshold for the duration of the timestep. To accurately simulate the effect of noise, the simulation timestep must be over $10^3$ times smaller than the noise cutoff frequency, making array-level simulations prohibitively expensive. 

To enable longer timesteps, we calculate at each timestep the probability of a threshold crossing since the previous timestep using first-passage-time theory.
We model shot noise as an Ornstein-Uhlenbeck process, which has the dynamics of first-order low-passed white noise~\cite{bibbona2008theornsteinuhlenbeck}. This assumption is valid in practice under the conditions that \Ipr is large compared to \Isf~\cite{graca2023optimalbiasing}, and that there is a clear dominant pole between \taupd and $\tausf=\frac{\Csf}{\gssf}$.
For an Ornstein-Uhlenbeck process, the probability of a threshold crossing between timesteps can be approximated in simulation using the method described in \cite{giraudo1999animprovedtechnique}. We compute two Bernoulli trials at each timestep to simulate if either the ON or OFF threshold were momentarily crossed between timesteps. In the case of a crossing, the crossing time is randomly sampled from the first-passage-time distribution~\cite{giraudo1999animprovedtechnique}.

\begin{figure*}[h]
    \centering
    \includegraphics[width=\textwidth]{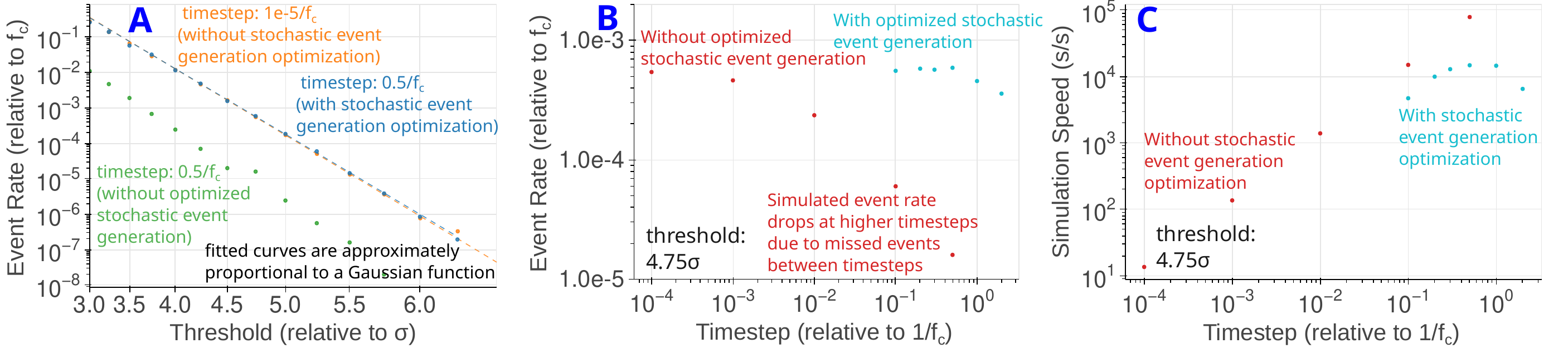}
    \caption{Noise event simulation with and without optimized stochastic event generation. (A) shows the event rate variation with changing the event threshold, (B) shows how the event rate at a fixed threshold decreases with increased timestep, and (C) shows how the simulation runtime increases with the timestep (simulation speed is calculated as the ratio between single-pixel simulated time and wall-clock runtime on a desktop computer Intel i7-8700K, running python 3.12.2 on Linux 6.13.6). $\text{f}_\text{c}$ denotes the noise cutoff frequency, and $\sigma$ denotes the noise standard deviation.}
    \label{fig:noise_events_sim_plots}
\end{figure*}

\cref{fig:noise_events_sim_plots} shows results from a simulation of noise events under constant background illumination, under the assumption that noise is first-order low-pass filtered white noise, following the same methodology as in \cite{graca2021unravelingtheparadox,graca2024eventcamera}.
\cref{fig:noise_events_sim_plots}\textbf{A} shows how optimizing stochastic event generation by predicting events between timesteps leads to accurate prediction of noise event rates even with timesteps of $\frac{0.5}{f_\text{c}}$, where $f_\text{c}$ is the noise cutoff frequency. Using the same timestep without any optimization (i.e. only assessing if the threshold was crossing at each timestep) leads to severe underestimate of noise event rates.
\cref{fig:noise_events_sim_plots}\textbf{B} shows how optimizing event generation allows increasing the simulation timestep by a factor of 3 orders of magnitude without losing accuracy.
\cref{fig:noise_events_sim_plots}\textbf{C} demonstrates wall-clock speedups of over 100x  without accuracy loss. 

\section{Conclusion}
\label{sec:conclusion}
We present two main contributions towards a physically realistic and computationally efficient DVS pixel model: 1. Stochastic event generation allows increasing the simulation timestep by 1000x without loss of accurate noise prediction, and 2. fitting a theoretical model derived from circuit analysis with physically meaningful parameters derived from bias currents, SPICE simulation, and pixel measurements leads to a model that can generalize over different illuminances at different biases, accurately predicting noise and signal response. These two results are important steps towards an accurate DVS model to be used for offline bias optimization and for the generation of physically accurate event datasets in challenging HDR conditions.
We intend to incorporate these improvements into the array-level simulator \href{https://sites.google.com/view/video2events/home}{v2e}~\cite{hu2021v2e}.

\subsection*{Acknowledgments}
This work was funded by the Swiss National Science Foundation SciDVS grant 185069 and Samsung Global Research.

\renewcommand*{\bibfont}{\scriptsize}
\footnotesize{\printbibliography}

\end{document}